\documentclass[aps,twocolumn,prl,showpacs]{revtex4}

\usepackage{graphpap}
\usepackage[dvips]{graphicx}
\usepackage[dvips]{graphics}
\usepackage{epic}
\usepackage{epsfig}    
\usepackage{color}

\newcommand{\expval}[1]{\left< #1 \right>}

\newcommand{\vf}{v_{\mathrm{F}}}
\newcommand{\pp}{n_i}
\newcommand{\hsgwl}{\beta}
\newcommand{\cc}{v_{\mathrm{F}}}

\begin{document}
%-----------------------------------------------
\title{Graphene Quantum Dots: Beyond a Dirac Billiard}
%-----------------------------------------------
 \author{Florian Libisch$^{1}$%\footnote{Corresponding author, e-mail:
                              %%florian@concord.itp.tuwien.ac.at}
                 , Christoph Stampfer$^{2}$, and Joachim Burgd\"orfer$^1$}
 \affiliation{$^1$Institute for Theoretical Physics, Vienna University of 
Technology\\Wiedner Hauptstra\ss e 8-10/136, A-1040 Vienna, Austria, European
Union \\
 $^2$Solid State Physics Laboratory, ETH Zurich, 8093 Zurich, Switzerland} 
\date{ \today}

\begin{abstract}
We present realistic simulations of quantum confinement effects in ballistic
graphene quantum dots with linear dimensions of 10 to 40~nm. We determine
wavefunctions and energy level statistics in the presence of disorder
resulting from edge roughness, charge impurities, or short-ranged
scatterers. Marked deviations from a simple Dirac billiard for massless fermions are
found. We find a remarkably stable dependence of the nearest-neighbor 
level spacing on edge roughness suggesting that the roughness of fabricated
devices can be potentially characterized by the distribution of measured Coulomb blockade peaks.

\end{abstract}

% Single particle states, Graphite/Carbon
% Quantum Chaos

\maketitle 

\begin{figure}
\epsfig{file=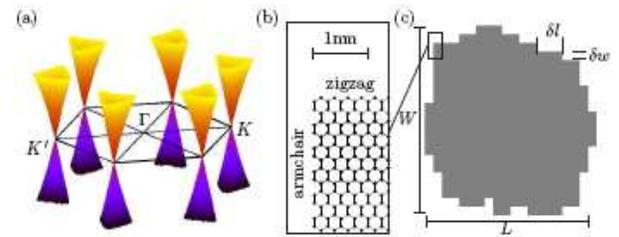,width=8cm}
\caption{(color online) (a) Graphene dispersion near $K$ and $K'$ points of the infinitely
  extended sheet. (b) Rectangular segment of a graphene flake, vertical edge armchair,
  horizontal edge zig-zag terminated. (c) (Approximately) rectangular quantum
  dot with rough edges.}
\label{fig_geometries}
\end{figure}

Graphene~\cite{nov05,zha05}, the first true two-dimensional (2D) solid, 
is attracting considerable attention, mostly due to unique dynamics of electrons
near the Fermi energy which closely mimics that of a massless Dirac
Hamiltonian [see Fig.\ref{fig_geometries}(a)]. Moreover, the double cone
structure near the $K$ and $K'$ points of the sublattices in reciprocal space
gives rise to a near ``pseudospin'' degeneracy, suggesting an analog of Dirac
four spinors. Envisioned applications range from high-mobility
nanoelectronics~\cite{gei07}, spin-qubits in graphene quantum
dots~\cite{tra07} and the creation of ``neutrino''
billiards~\cite{ber87,pon08}. Dirac (including neutrino) billiards
receive growing interest as a complement to classical and quantum
(Schr\"odinger) billiards, which have taken central stage in studies
elucidating the quantum-to-classical crossover in both regular and chaotic devices.
Additionally, spin coherence times in graphene are expected
to be very long due to weak spin-orbit and hyperfine
couplings~\cite{min06,tom07} making graphene quantum dots promising for future
spin based quantum computation~\cite{tra07}.  However, confining electrons in
graphene is a challenge, due to the gap-less electronic structure and
the Klein tunneling paradox~\cite{dom99,kat06}.  This difficulty has recently
been overcome by structuring 2D graphene and quantum mechanical confinement
effects have been observed in nanoribbons~\cite{che07,han07,lin08},
interference devices~\cite{mia07}, single electron
transistors~\cite{sta08,sta08b} % [nov07,sta08,sta08b] 
and graphene quantum billiards~\cite{pon08}.

In the following we present a realistic simulation for the single particle
spectrum of graphene quantum dots (i.e. billiards) by explicitly considering
rough edges and disorder. This work was motivated by recent advances in
fabricating dots with linear dimension $d$ ranging from a few hundred nm down
to about 40~nm, and determining their nearest neighbor energy level spacing
distribution~\cite{pon08,sta08b}. We analyze dot wavefunctions, the density
of states (DOS) and the nearest neighbor spacing distribution (NNSD). We
address the question to what extent the electron spectra now experimentally
accessible via measurements of Coulomb blockade peaks reveal information on
the roughness and size of the graphene quantum dot. To put it provocatively:
Can one ``hear'' the rugged shape of a drum if it is made of a graphene flake?

We investigate graphene dots with linear dimensions between 10 and 40~nm,
%containing between 6000 and 75000 carbon atoms. This size agrees 
in line with
currently fabricated devices~\cite{pon08}. The shape is chosen regular in
the absence of edge roughness. One motivation of this
choice was the remarkable result~\cite{ber87} that a Dirac neutrino billiard,
in sharp contrast to a Schr\"odinger billiard, would feature chaotic dynamics.
We consider a ballistic dot, thereby neglecting inelastic scattering.  
%inside the dot
This is justified as the inelastic mean free path $\lambda_{il}$
found in experiment exceeds the linear dimension $d = \sqrt{4LW/\pi}$ of the
dot, $d \le 40$nm $\ll \lambda_{el} \approx400$~nm~\cite{nov05}. Our
simulation allows for the inclusion of disorder through (i) edge roughness
[see Fig.~\ref{fig_geometries}(c)], (ii) short-range disorder due to 
%randomly distributed 
point defects in the interior, and (iii) long-range screened
Coulomb distortion due to charge deposition.
% in either the substrate or the flake. 
Rough edges are simulated by modulating the boundary of the dot by
steps of height $\pm\delta w$ and length $\pm\delta l$ randomly chosen from
the interval $[0,\Delta W], \;\Delta W \ll d$.  We refer to $\Delta W$ as the
amplitude of edge roughness which varies between 0.3~nm (weak disorder) and
2~nm (strong disorder). The resulting piece-wise straight edge features
alternating zig-zag and armchair sections [see
Figs.~\ref{fig_geometries}(b,c)].  We describe impurities and defects in the
flake by positioning either long range ($V(\mathbf{r}) = V_0 e^{ -\alpha
|\mathbf{r}-\mathbf{r}_0|}$) or short range ($V(\mathbf{r}) =
\delta(\mathbf{r}-\mathbf{r}_0)$) scatterers at randomly selected lattice
sites $\mathbf{r}_0$. We use an impurity density $\pp < 1.8\cdot 10^{-3}$
impurities/carbon (10 to 100 defects per flake), as estimated by recent 
work~\cite{adam}.

The spectrum of the graphene quantum dots is
determined employing a Lanczos algorithm~\cite{Lanczos} giving the 500
eigenstates closest to the Fermi edge.  The graphene flake is described by a
third nearest neighbor tight-binding approximation to correctly reproduce the
graphene bandstructure \cite{TBReich}. The modified C-C bond length at the
flake boundary is accounted for by increasing nearest-neighbor coupling to the
outmost carbon atoms by 12\% in accordance with recent ab-initio density
functional calculations~\cite{Louie}.  Our ensemble averages for the DOS
$\expval{\rho}_\xi = \langle\sum_i\delta(E-E_i)\rangle_{\xi}$ encompass
typically 5000 disorder realizations $\xi$.

\begin{figure}[h]
\epsfig{file=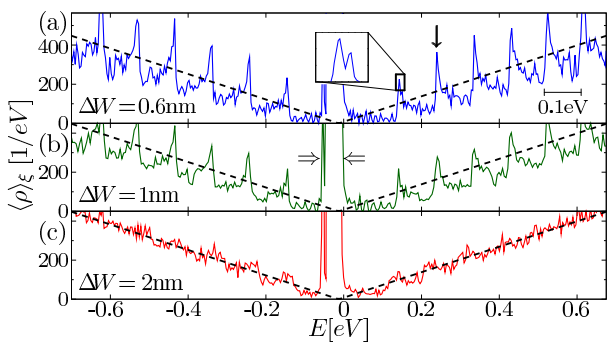,width=8cm}
\caption{(color online) Ensemble-averaged density of states
  $\expval{\rho(E)}_\xi$ of graphene quantum dots with increasing edge
  roughness, see different values for $\Delta W$ in the subfigures.  The size
  of all devices is equal, $d = 20 nm$ (20.000 atoms), their width $W = 16nm$.
  Dashed lines indicate the averaged linear DOS for Dirac billiards [see
  Eq.~(\ref{eq:DOS})]. The inset shows the $K-K'$ splitting of $12meV$.}
\label{fig_DOS}
\end{figure}

The linear dispersion relation of a massless Dirac particle implies a DOS
  linear in $\varepsilon$,
\begin{equation}
\rho(\varepsilon) =  \frac{1}{2(\hbar \vf)^2} d^2 \left|\varepsilon\right|, \label{eq:DOS}
\end{equation} 
where $d=\sqrt{4 WL/\pi}$ is the effective diameter of a dot with area $WL$ and
$\varepsilon$ is measured relative to the conical intersection
[Fig.~\ref{fig_geometries}(a)] assumed to coincide with the Fermi edge. The
simulated DOS for the quantum dots display marked deviations from
Eq.~(\ref{eq:DOS}). For weak disorder pronounced size quantization peaks
appear [see e.g. vertical arrow in Fig.~\ref{fig_DOS}(a)].  Their positions
are determined by the smallest linear dimension of the flake~\cite{sizequ}.
We have investigated both cases $W < L$ ($W > L$), and find the same
qualitative behavior. In the following, we assume $W < L$. Note that width
($W$) and length ($L$) of the rectangular flake are not strictly equivalent as
the vertical boundary features an arm-chair border while the horizontal forms
a zig-zag border. The distance between the quantization peaks is $\Delta E =
\hbar \vf \pi/ W \approx 1.5/ W [eV]$, width $W$ given in nm.  This yields
an analytic prediction for the energy separation of $0.1 eV$ for the peaks in
Fig.~\ref{fig_DOS}. Weak disorder, i.e.~small edge roughness can induce
coupling between the cones at $K$ and $K'$ [Fig.~\ref{fig_geometries}(a)].
This manifests itself by a fine structure of size quantization by
lifting the degeneracy [inset in Fig.~\ref{fig_DOS}(a)]. The quantum
confinement peaks in the graphene dot are enhanced compared to a corresponding
Schr\"odinger billiard of the same geometry in part because of both the
altered dispersion relation and the additional degeneracy. Strong disorder
smears out size quantization patterns and the DOS begins to resemble that
of a zero-mass Dirac fermion in free space (Eq.~\ref{eq:DOS}). Only when the
edge roughness can be limited to the sub-nanometer scale, quantized
conductance in graphene nanoribbons persists~\cite{lin08}. 

\begin{figure}[t]
\epsfig{file=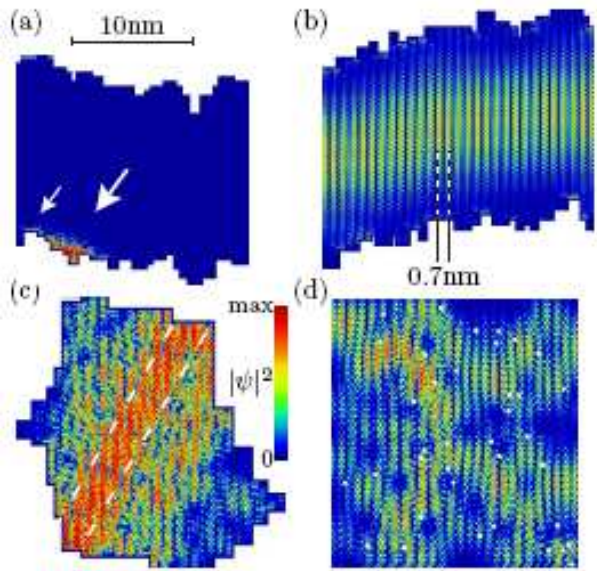,width=8cm}
\caption{(color online) Eigenstates of graphene quantum dots with dot area
of about $225 nm^2$ (13.000 carbon atoms). Eigenenergies are
(a) 65 meV (b) 220 meV (c) 650 meV (d) 1020 meV respectively. The
edge roughness parameter $\Delta W = 1nm$ in (a-c) while 
$\pp = 0.005$ in (d) (White dots mark the center of long-range impurities).}
\label{fig_wave}
\end{figure}

Even in the limit of strong disorder, the prominent peak in the DOS near the
Fermi edge remains unchanged [see horizontal arrows in Fig.~\ref{fig_DOS}(b)].
A direct look at the wavefunction [Fig.~\ref{fig_wave}(a)] reveals its origin:
A large number of strongly (Anderson-)localized states at the edges of the
graphene flake. Each eigenstate features a non-vanishing amplitude only at a
few, not always spatially connected carbon atoms, with a decay length into the
bulk of typically $0.5$~nm [see left arrow in Fig.~3(a)]. We find that the
eigenenergies of these states are extremely sensitive to the site energies at
the corresponding lattice sites.  This agrees well with the experimental
observations of sharp resonance in the electron transport through graphene
constrictions~\cite{sta08}. 
%We expect that different functional groups
%(e.g.~H- or OH-) attached to the outermost carbon atoms may strongly influence
%the local DOS and the position of this localization peak relative to the Fermi
%edge.

Delocalized states contributing to the size quantization peaks show pronounced
features well beyond the simple picture of a confined zero-mass Dirac
particle. While the transverse quantization resembles that of a conventional
conductor, the interference pattern in the electron probability density
[Fig.~\ref{fig_wave}(b)] results from the simultaneous presence of multiple
wavelength scales for the cone near the $K$ point (unlike wavefunctions near
the $\Gamma$ point), in $\mathbf{k}=(k_0+\cc\pi/L,\cc\pi/W)$: Parallel to
armchair edges (i.e.~in vertical direction in Fig.~\ref{fig_wave}), the
wavelength is of the order of twice the width of the ribbon-like dot $\approx
32$~nm.  Parallel to zigzag edges, (i.e.~in horizontal direction in
Fig.~\ref{fig_wave}) the wave oscillations are much shorter with a typical
wavelength of $0.7$ nm [see Fig.~\ref{fig_wave}(b)] resulting from beating
(frequency ratio 3:2) between lattice periodicity $a =0.24$~nm and the
characteristic wavelength $\lambda_0 = 2\pi/k_0 \approx 0.37$~nm, where
$k_0\gg \cc\pi/L$ is the distance between $\Gamma$ and $K$ point in reciprocal
space [see Fig.~1(a)]. We find beating patterns with this characteristic
length scale to be universally present in all delocalized states, even in the
presence of long-range disorder. Only because of the sub-nanometer length
scale of $\lambda_0$ at the $K$ point is the graphene dot sensitive to edge
roughness and disorder on a length scale of a few nanometers, in contrast to a
Dirac cone at the $\Gamma$ point. The reason for the $K$-point $(k_0,0)$ [as opposed 
to those at 60 and 120 degrees, $\frac 12(k_0,\pm \sqrt 3 k_0)$], to
appear in the eigenstate shown is the orientation of the zigzag (armchair)
edges in the flake parallel (orthogonal) to the $(k_0,0)$ direction. %Note,
However, all three directions appear for higher transverse quantum
numbers, resulting in enhancements along the three zigzag-directions of the
lattice [i.e.~horizontal, 60 and 120 degrees, see dashed lines in
Fig.~\ref{fig_wave}(c)]. As a consequence, eigenstates feature a 2D hole
(``swiss-cheese'') pattern emerging from the interference of plane waves
rotated by 60 degrees relative to each other [see Fig.~\ref{fig_wave}(d)].

\begin{figure}[h]
\epsfig{file=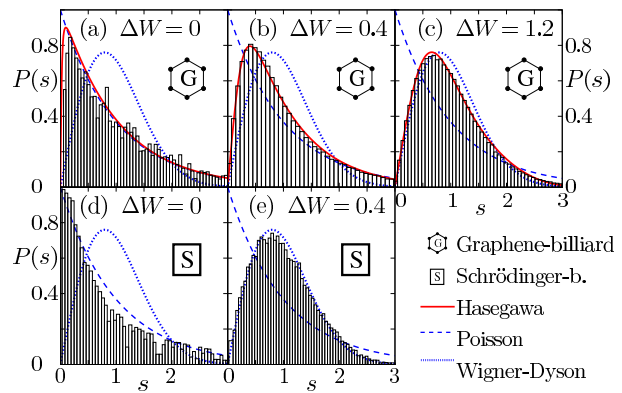,width=8cm}
\caption{(color online) NNSD $P(s)$ of  
  different billiards: Rectangular graphene flake with (a) smooth edges
  ($\hsgwl = 0.07$).  (b) finite edge roughness $\Delta W
  = 0.5$~nm$\;(\hsgwl = 0.4)$, (c) $\Delta W = 1$~nm$\;(\hsgwl = 2.5)$.
  (d,e) Schr\"odinger billiards with same edge roughness as (a,b).
  The solid red curve shows fits to the Hasegawa distribution.  Dashed
  (dotted) lines show a Wigner-Dyson (Poisson) statistic as guide to the eye.
}
\label{fig_dists}
\end{figure}

In order to delineate the influence of disorder and edge roughness on the
energy level statistics, we have determined the NNSD, $P(\Delta E)$, i.e.~the
probability that the energy difference between two adjacent energy levels is
$\Delta E$, for different amplitudes $\Delta W$ of roughness. Within the
framework of quantum dynamics of Schr\"odinger billiards, $P(\Delta E)$
follows a Poisson distribution for separable (classically regular) shapes
while it should display a Wigner-Dyson (or GOE) distribution for irregularly
shaped (classically chaotic) billiards. In contrast, even rectangular shaped
Dirac neutrino billiards have been shown to feature a GUE distribution because
of the broken time reversal symmetry due to chirality~\cite{ber87}. After spectral
unfolding [$s= \overline\rho(\varepsilon_i)$, with $\expval{P(s)}=1$ and
$\expval{s P(s)}=1$] we find for the ideal rectangular graphene dot ($\Delta W
= 0$) a near-perfect Poisson distribution [see Fig.~\ref{fig_dists}(a)].  By
gradually increasing either the edge roughness or the defect concentration
$\pp$, the distribution smoothly evolves into a Wigner-Dyson like statistics
[see Fig.~\ref{fig_dists}(b,c)]. Clearly, such a behavior reflects the
conservation of time-reversal symmetry in graphene quantum dots. Among the
distribution functions suggested for the transition regime for classically
mixed phase space \cite{Brody,Izraelev,Caurier,Haake,Berry}, the best
fit for the disorder parameters and geometries investigated was achieved for
the two-parameter Hasegawa distribution~\cite{hasegawa}
\begin{equation}
P_H(s;\alpha,\hsgwl) = N \frac{\rho s e^{-\rho s - (\alpha \rho s)^2/2}}
{\sqrt{\rho^2 s^2 e^{-\alpha^2 \rho^2 s^2} + \hsgwl^2
  e^{-2 \rho s}}},\label{eq:hasegawa}
\end{equation}
where $\rho$ and $N$ are determined by the normalization conditions $\expval
{P_H} = \expval{s P_H}$ = 1 \cite{hasegawa}. While the control parameter
$\hsgwl$ describes the transition from Poissonian ($\hsgwl=0$) to Wigner-Dyson
statistics ($\hsgwl \rightarrow \infty$), $\alpha$ is a system-specific
constant. Indeed, we find $\alpha = 0.75$ to correctly reproduce our
numerically obtained NNSD for different values of both edge roughness as well
as scatterers [see Fig.~\ref{fig_dists}(a-c)].  A strong edge roughness of $2$nm
(or impurity concentration $\pp = 5\cdot 10^{-3}$) is required to reach the completely chaotic limit,
i.e.~a Wigner-Dyson NNSD statistics.  Remarkably, for moderate values of the
edge roughness amplitude ($\Delta W =0.6$~nm) a Schr\"odinger billiard and a
graphene billiard of the same geometry display a markedly different NNSD
(Fig.~\ref{fig_dists}): While the Schr\"odinger billiard has already reached
the Wigner-Dyson limit ($\hsgwl \rightarrow\infty$), for the graphene the NNSD
still is closer to the Poisson limit, pointing to the unique spectral
properties of graphene. The quasi-regular dynamics in graphene is
more stable against disorder than in corresponding Schr\"odinger
billiards. The origin of this, at first glance, surprising finding is closely
related to the electronic structure of graphene at the $K$ point
[Fig.~\ref{fig_geometries}(a)]. In a classical rectangular ballistic billiard with
only rectangular edges along the armchair or zig-zag direction
%(Fig.~\ref{fig_geometries} and \ref{fig_dists}) 
 an additional constant of motion,
the magnitude of the linear momentum $\left|\mathbf{k}\right|$, exists. Such
billiards are therefore classically regular irrespective of the number or size
of the edges. By contrast, due to the larger deBroglie wavelength of 
Schr\"odinger billiards with wavenumbers near the $\Gamma$ point
cannot resolve the exactly rectangularly shaped edges, and thus mimics
chaotic dynamics. 
%For the same size of the edges $\Delta W$, the graphene
%eigenstate features a much shorter wavelength due to the position of the Dirac
%cone near the $K$ point in the Brillouin zone. The quantum dynamics of the
%graphene billiard is therefore closer to the classical limit and its level
%statistics closer to the Poisson limit of regular dynamics.

\begin{figure}
  \epsfig{file=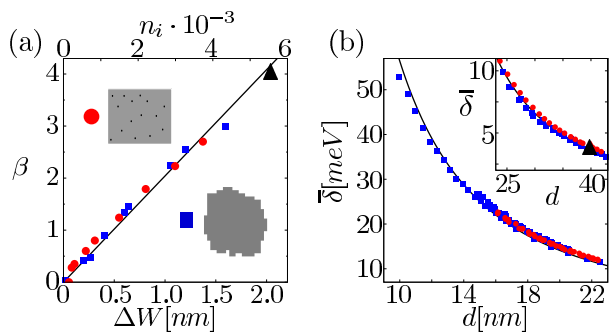,width=8cm}
\caption{(color online) (a) Dependence of the control parameter $\beta$ for the
  transition from a Poisson to a Wigner-Dyson distribution
  (Eq.~\ref{eq:hasegawa}) on the edge roughness amplitude $\Delta W$ or the
  defect density $\pp$. (b) Dependence of rescaled mean-level spacing
  $\overline\delta=\expval{\Delta E \cdot E/E_0} $meV, where $E_0$ was taken
  at $100$meV, on the size of the quantum dot. Triangles represent
  experimental data, rescaled by $\Delta E \rightarrow (\Delta E - E_C)\frac
  {E}{E_0} = (\Delta E - 5)\frac{50}{100}$ to take into account the charging energy $E_C$ and the
  energy dependence of the DOS.  \cite{pon08}.}
\label{fig_var}
\end{figure}

Of potential technological significance is the dependence of the NNSD on the
disorder in graphene billiards. % which might be used as a quantitative indicator.
For all three classes of disorder we consider (edge roughness, short and long
range disorder) we find a linear relation between the NNSD parameter $\hsgwl$
and the edge roughness amplitude, $\hsgwl \approx 2\Delta W$, and between
$\hsgwl$ and the defect density, $\hsgwl \approx 0.7 \pp$
[Fig.~\ref{fig_var}(a)]. 
As $\hsgwl$ can be obtained with high precision from a fit to $P(s)$, the edge
roughness or defect density can be deduced from the NNSD, if the distribution
is Poisson-like in the limit $\Delta W=0$ ($\pp = 0$). We suggest that this
dependence could be used to estimate the disorder in experimentally realized
regularly shaped ballistic graphene quantum dots.  Numerically, we find
$\hsgwl \approx 4$ in a fit to data from recent experimental investigations of
a $40$nm graphene billiard \cite{pon08}, corresponding to an effective
roughness $\Delta W_e \approx 2$~nm, or an effective defect rate $\pp
\approx 5.5\cdot 10^{-3}$ (see black triangles in Fig.~\ref{fig_var}).  
As the
second moment $\sigma = \expval{s^2 P(s)}_\xi$ of the NNSD decreases with
increasing disorder, one could alternatively obtain an estimate for the roughness
of a flake from $\sigma$.  We find however, that the dependence on $\hsgwl$ is
numerically more reliable, as the entire distribution is used for a fit to
$P_H$.

While the shape of the
unfolded NNSD, $P(s)$, is sensitively dependent on disorder, it is to a good
degree of approximation size independent, as our numerical data confirms (not
shown). The reason for this is the normalization $\expval{s P(s)} = 1$, which
scales out size-related effects. 
By contrast, the absolute level
spacing $\expval{\Delta E}$ contains direct information on the size of quantum
confinement. By rescaling each level spacing by the local energy, relative to
a fixed energy $E_0 = 100meV$, one obtains the energy independent
expectation value 
\begin{equation}
%$
\overline\delta = \expval{\Delta E\cdot\frac{E}{E_0}} = 
(\hbar \vf)^2\frac{2}{d^2 E_0} = \frac{5500\,nm^2}{d^2} meV.%$
%\label{eq:delta} 
\end{equation}
This rescaled mean level spacing is, indeed, independent of edge roughness and
disorder [see Fig.~\ref{fig_var}(b)]. %Note that in spite of the large
%contribution of Anderson localized states near the Fermi edge the mean level
%spacing accurately follows Eq.~(\ref{eq:delta}). 
%To achieve agreement with
%Eq.(\ref{eq:delta}) we have included in the ensemble the states up to 1eV away
%from the Fermi edge. The increased spacing of the more distant levels offsets
%the clustering of the localized levels near the Fermi edge. 
Agreement with the
experimental data \cite{pon08} for 
% point near 
$d = 40$~nm is surprisingly good.

In conclusion, the spectrum of realistic graphene quantum
dots in the presence of disorder (edge roughness or defects) reveal unique
features which differ from both Schr\"odinger or Dirac billiards of confined
massive or massless free particles. The graphene bandstructure near the $K$
point leaves clear imprints. They include interference structures in the
wavefunctions, enhanced confinement effects, and a delayed transition from
Poisson to Wigner-Dyson nearest-neighbor distributions. While one still
``cannot hear the imperfect shape of the drum'', the size and roughness of graphene
quantum dots can be, indeed, inferred from the spectral properties.

\begin{acknowledgments}
We thank K.~Ensslin, T.~Ihn, S.~Rotter and L.~Wirtz
for valuable discussions.
F.L. and J.B. acknowledge support from the FWF-SFB Adlis, 
C.S. support from NCCR. 
\end{acknowledgments}

\end{document}